\documentclass[aps,prl,twocolumn,superscriptaddress,showpacs,floatfix]{revtex4}

\usepackage{graphicx}
\usepackage{dcolumn}
\usepackage{bm}
\usepackage{color}%

\begin{document}

\title{\hfill {\small Phys. Rev. Lett. {\bf 108}, 107205 (2012)}\\
       Magnetism and anti-ferroelectricity in MgB$_6$}

\author{Igor Popov}
\author{Nadjib Baadji}
\author{Stefano Sanvito}
\affiliation{School of physics and CRANN, Trinity College
             Dublin 2, Ireland}

\date{\today}


\begin{abstract}

We report on a density functional theory study demonstrating the coexistence of weak ferromagnetism and anti-ferroelectricity in boron-deficient MgB$_6$. 
A boron vacancy produces an almost one dimensional extended molecular orbital, which is responsible for the magnetic moment formation. Then, long range 
magnetic order can emerge from the overlap of such orbitals above percolation threshold. Although there is a finite density of states at the Fermi level, 
the localized nature of the charge density causes an inefficient electron screening. We find that the Mg$^{2+}$ ions can displace from the center of 
their cubic cage, thus generating electrical dipoles. In the ground state these order in an anti-ferroelectric configuration. If proved 
experimentally, this will be the first material without $d$ or $f$ electrons displaying the coexistence of magnetic and electric order. 

\end{abstract}



\maketitle


The discovery of weak ferromagnetism in La-doped CaB$_6$ by Young \textit{et al.}~\cite{Young} ignited a surge of research on 
alkaline-earth hexaborides, XB$_6$ (X = Ca, Sr, Ba)~\cite{Tromp,Monnier,Edwards,Lee_APL,Okatov,Dorneles,Cho}. Initially 
the origin of the magnetism in these materials was attributed to possible transition-metal impurities~\cite{Matsubayashi}. However, 
subsequent experiments in high-purity CaB$_6$, SrB$_6$ and BaB$_6$ crystals~\cite{Vonlanthen,Ott,Lofland} clearly indicated 
that the magnetism may be an \textit{intrinsic} property of the XB$_6$ family, and not due to contaminations. Theoretical predictions 
for defective XB$_6$ assigned to boron vacancies, V$_\mathrm{B}$, the primary cause of the magnetism. These introduce flat bands 
near the Fermi level, which spin-split due to the strong exchange interaction in the B $p$-shell~\cite{Janak,Monnier,Cao,Edwards}. 
Such predictions have been recently supported by experimental evidence~\cite{Cho,Maiti_2007,Maiti_2008}, which also extends to 
CaB$_6$ and SrB$_6$ thin films~\cite{Dorneles}.

In contrast to the heavy compounds of the XB$_6$ family, MgB$_6$ has received only little attention. This is mainly due
to an early theoretical calculation, performed with empirical methods, suggesting the poor stability of MgB$_6$  compared to 
the more thermodynamically stable MgB$_2$, MgB$_4$, MgB$_7$ and MgB$_1$$_2$~\cite{Ivanovskii_InorgMat}. In addition
the small Mg ionic radius, second only to Be among the alkalines, may result in a high ionic mobility and thus in Mg segregation. 
However, in contrast to this early prediction MgB$_6$ has been often observed experimentally \cite{MgB6exist}, most notably  
with conclusive high resolution tunnelling electron microscopy \cite{Li_APL}. In this letter we report on a density functional 
theory (DFT) study of the structural, electronic and magnetic properties of V$_\mathrm{B}$-rich MgB$_6$. Our results point to a 
ferromagnetic ground state, which coexists with an anti-ferroelectric dipolar order. As such MgB$_6$ appears to be the first 
example displaying the coexistence of magnetic and electric order in a material not containing ions with partially occupied $d$ or 
$f$ shells. 

Calculations are performed with the DFT numerical implementation contained in the {\sc Siesta} code~\cite{Siesta}. The 
Perdew-Burke-Ernzerhof form of the generalized gradient approximation~\cite{PBE} is used for the structural optimization and 
for the calculation 
of the phonon dispersion. Additional calculations of the electronic properties use the Perdew-Zunger local density approximation 
(LDA)~\cite{PerdewZunger} including atomic self-interaction corrections (ASIC)~\cite{Das_ASIC}. The core electrons are described 
by norm-conserving pseudopotentials with partial core corrections~\cite{TroullierMartins}, while the wave-function is expanded 
over a double-zeta numerical atomic orbital basis set including the following orbitals: Mg 3$s$, Mg 3$p$, B 2$s$ and B 2$p$. 
The charge density and the DFT potential are determined over a real-space grid with an equivalent mesh cutoff energy of 350~Ry, 
but a value of 700~Ry is used for calculating the phonon spectra. A 1728 k-points dense mesh samples the Brillouin zone. 
Atomic relaxation is performed by conjugate gradient until the forces are smaller than 0.04 eV/\AA\ (0.001 eV/\AA~for the phonon 
spectrum). Phonons are calculated by using the elementary unit cell, hence only optical modes are considered.

\begin{figure}
\includegraphics[width=\columnwidth]{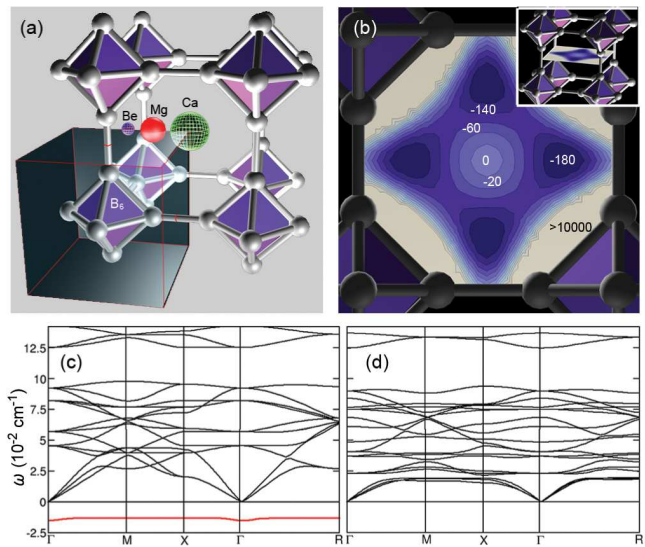}
\caption{\label{inversionInstability} (colour online) MgB$_6$ structural properties. In (a) the XB$_6$ (X=Be, Mg, Ca) crystal
structure, where the gray volume includes the unit cell. Note the different displacement of Ca, Mg and Be within the 
B$_6$-defined cubic cage. (b) Potential energy surface (isovalues are in units of meV) for Mg$^{2+}$ across the median plane 
of the cubic cage (shown in the inset). Phonon band structure for the center-symmetric MgB$_6$ crystal (c) 
and center-symmetric CaB$_6$ crystal (d).}
\end{figure}
We start our analysis by first looking at the defect-free crystal. XB$_6$ (X=Be, Mg, Ca) crystallizes in a halite cubic structure with the 
two inequivalent sites occupied respectively by the divalent X$^{2+}$ ion and a B$_6^{2-}$ octahedral cluster [see Fig.~\ref{inversionInstability}(a)]. 
The lattice parameter is determined by the size of the B$_6$ clusters and it is calculated to be 4.115~\AA~for MgB$_6$. The position of the 
cation however depends on the specific element. After structural relaxation Ca remains in the center-symmetric position, while Mg moves 
towards the face of the cubic cage and nests at 1.143 \AA~away from it. Interestingly, Be moves completely into the B$_6$ plane. 

The potential energy surface of Mg$^{2+}$ is plotted in Fig.~\ref{inversionInstability}(b) over one of the median planes cutting
across the cubic cage, with the zero energy taken at the center-symmetric position. We observe four equivalent minima in the 
plane, which translate into six minima over the cubic volume. The potential barriers between the minima are around 70~meV, 
which indicate that the displaced Mg positions are stable at room temperature. Additional evidence of the Mg displacement is 
provided by the MgB$_6$ phonon spectrum [Fig.~\ref{inversionInstability}(c)] for the center-symmetric structure, which presents 
a triply-degenerate negative frequency mode. Note that the same is absent for the CaB$_6$ crystal 
[Fig.~\ref{inversionInstability}(d)], which remains undistorted.

\begin{figure}
\includegraphics[width=\columnwidth]{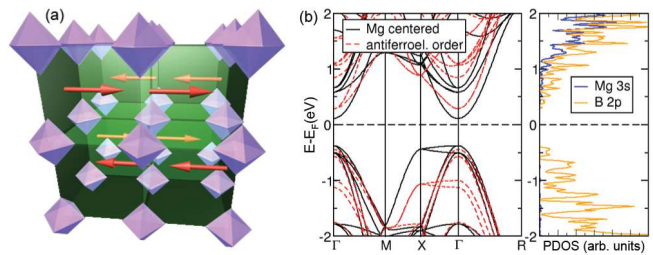}
\caption{\label{afe} (colour online) Electronic structure of defect-free MgB$_6$. In (a) we present the type-C antiferroelectric 
order for a $3\times3\times3$ supercell. The arrows indicate the direction of the electric dipoles (atoms and bonds are removed 
for simplicity). In (b) a comparison of the electronic band structures for the center-symmetric and the type-C anti-ferroelectric 
ordered system obtained for $2\times 2 \times 2$ supercell. The corresponding PDOS is presented in the rightmost panel.}
\end{figure}

Given the ionic character of MgB$_6$, the displacement of Mg$^{2+}$ generates an electrical dipole parallel to the direction of the 
displacement. Standard (LDA) Berry-phase 
calculations give us an electrical polarization of 57.6~$\mu$C/cm$^2$ for an hypothetical ferroelectric phase. This however does not 
represent the correct long-range ferroic order. The overall dipolar arrangement is then computed by comparing total energy 
calculations for a $2\times2\times2$ supercell with different dipole configurations (ferroelectric, antiferroelectric of type A and
type C and ferroelectric with only one dipole chain rotated)~\cite{AFEtypes}. We find the energy minimum at the type-C anti-ferroelectric 
arrangement, which is illustrated in Fig.~\ref{afe}(a). Such a state is 226~meV lower than the ferroelectric one (per elementary 
unit cell), which suggests that it might be a relatively high-temperature phase.

The projected density of states (PDOS) of MgB$_6$ indicates that the material is semiconducting with a relatively small
indirect (LDA) band gap of about 0.5~eV [see~Fig.~\ref{afe}(b)]. This reduces to 0.27~eV if the calculation is performed at the GGA level. 
The conduction and valence bands have respectively Mg-$s$ and B-$p$ character, 
as expected from its ionic Mg$^{2+}$(B$_6$)$^{2-}$ nature. An ionic bonding favours the Mg$^{2+}$ ion to move off-center in 
order to minimize the electrostatic energy. The same mechanism is ineffective for the heavier CaB$_6$ simply because of the 
much larger ionic radius of Ca$^{2+}$. Furthermore, as there is little covalent bonding between Mg and B$_6$ also the interaction 
between the dipoles has electrostatic nature and so an anti-ferroelectric order is expected~\cite{Nicola_Review}. Finally we note 
that the Mg displacement has little effect on the band structure as the anti-ferroelectric ground state remains insulating with 
an LDA gap of 0.73~eV [Fig.~\ref{afe}(b)].

\begin{figure}
\includegraphics[width=\columnwidth]{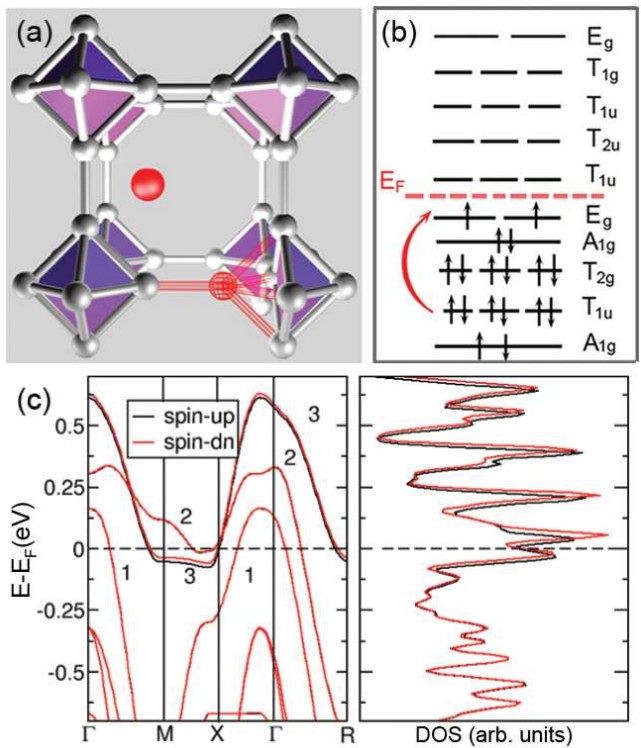}
\caption{\label{magnetism} (colour online) Electronic structure of V$_\mathrm{B}$-rich MgB$_6$. In (a) the 
geometry of a supercell containing one V$_\mathrm{B}$. (b) shows the energy state diagram of a free-standing B$_6$ cluster. 
The red arrow indicates the re-arrangement of states upon introduction of a vacancy in MgB$_6$ crystal.
The electronic band-structure and DOS for the cell containing one vacancy are in panel (c).}
\end{figure}

Having characterized the electronic structure of defect-free MgB$_6$, we now move to discuss the effect of V$_\mathrm{B}$ in the
crystal. The MgB$_{6}$ crystal bares symmorphic symmetry, so that its bands can be directly related to the energy levels of 
the B$_6$ cluster. The molecular states [Fig.~\ref{magnetism}(b)] are characterized according to the irreducible representations of 
the O$_h$ group. The low lying T$_{1u}$ state is of special interest. In fact a vacancy in B$_6$ crystal moves the T$_{1u}$-derived 
level upwards into the pristine gap of the crystal. T$_{1u}$ is threefold degenerate with each of the levels in the manifold dominated 
by only one of B-$p_i$ ($i=x,y,z$) orbitals. A vacancy also removes an electron so that B$_5$ is left with a single electron in non-degenerate 
orbital level. We then expect magnetic moment formation \cite{Das_Hf} with the spin-polarized charge density distributing along the 
direction where the B atom has been removed.  

This qualitative picture is further confirmed with numerical DFT calculations. Since the self-interaction error in LDA usually 
precludes the correct description of defects in semiconductors, here we employ the ASIC scheme~\cite{Das_ASIC}, 
which has been already proved successful in describing anion defects in the closely related MgO insulator~\cite{Droghetti_ASIC}. 
In particular we perform calculations by using different values of the screening parameter $\alpha$, ranging from $\alpha$=0 (LDA) 
to $\alpha$=1 (full ASIC), and compare spin-polarized (SP) and non-spin-polarized (NSP) results. 

Let us first discuss the magnetic moment formation associated to a single V$_\mathrm{B}$ in one $2\times2\times2$ 
supercell [see Fig.~\ref{magnetism}(a)]. LDA yields a cell magnetic moment of 3~m$\mu_\mathrm{B}$.
Importantly such a moment increases in the ASIC calculations. In fact we find 414~m$\mu_\mathrm{B}$, 509~m$\mu_\mathrm{B}$, 565~m$\mu_\mathrm{B}$ 
and 617~m$\mu_\mathrm{B}$ respectively for $\alpha=$~0.25, 0.5, 0.75 and 1. Also the total energy difference between the SP and the NSP 
solutions, $\Delta E(\alpha)=E_\mathrm{SP}-E_\mathrm{NSP}$, varies with the degree of ASIC considered ($\alpha$). In particular we find 
$\Delta E$=~2~meV for LDA and -23~meV, -44~meV, -116~meV and -296~meV respectively for ASIC with $\alpha=$~0.25, 0.5, 0.75 and 1. 
As such we find that both the magnetic moment and its energy stability increase with increasing the amount of self-interaction corrections
included. A value $\alpha=0.5$ is usually considered appropriate for semiconductors so that we estimate a magnetic moment
of around 1/2~$\mu_\mathrm{B}$, with a formation energy well above the room temperature. 


A deeper insight into the origin of the magnetic moment is obtained by looking at the band structure of the supercell containing 
one V$_\mathrm{B}$ [Fig.~\ref{magnetism}(c)]. As expected from the discussion about B$_6$ now the system has a metallic 
electronic structure with three bands crossing the Fermi level, $E_\mathrm{F}$ [labelled with 1, 2, and 3 in Fig.~\ref{magnetism}(c)]. 
Their dispersion is typical of a semi-metal with both electron (from M to X and at R) and hole (at $\Gamma$) pockets at the $E_\mathrm{F}$. As the three bands are 
non-orbital-degenerate the system is not prone to the Jahn-Teller distortion, which might compete with the formation of the 
moment. Importantly only the electron carrying band [3 in Fig.~\ref{magnetism}(c)] spin-splits, while the other two remain 
non-spin-polarized. Recalling the fact that the Stoner exchange parameter is rather large in the 2$p$ shell~\cite{Janak}, 
the spin-splitting occurs in bands with a narrow band-width, i.e. in bands that can provide a large DOS at $E_\mathrm{F}$. 
This is the case of the electron carrying band (3), which has a band-width of 0.68~eV to be compared with 
1.3~eV and 1.2~eV for the other two. Such a picture of magnetic moment formation due to narrow bands is robust with both the 
number and the position of the vacancies as we tested for a $2\times2\times2$ supercell with two V$_\mathrm{B}$'s at different 
positions.

\begin{figure}
\includegraphics[width=\columnwidth]{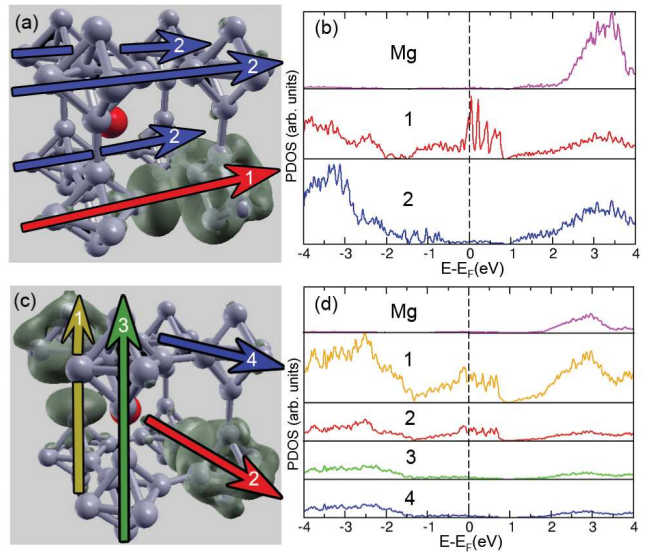}
\caption{\label{screening} (colour online) Local density of states around the Fermi level of a MgB$_6$ crystal with one (a) 
and two (c) vacancies per $2\times2\times2$ supercell. In the panels (b) and (d) we show the DOS projected over different 
sub-set of B atoms. In particular we project over B atoms along the directions indicated by the arrows in the panels (a) and (c).}
\end{figure}

The unique directional nature of the V$_\mathrm{B}$-derived electronic bands at $E_\mathrm{F}$ is provided in 
Fig.~\ref{screening}(a), where we show the local density of states (LDOS), obtained within the energy range 
($E_\mathrm{F}$-0.05eV, $E_\mathrm{F}$+0.05eV). From the figure it emerges that the electron cloud is delocalized only along 
the direction defined by the center of vacant B$_5$ cluster and by the position of 
V$_{B}$, as expected from the T$_{1u}$ symmetry. The contribution to the DOS from the B atoms which do not lie along such 
a preferential direction is negligible [Fig.~\ref{screening}(b)].
The presence of multiple V$_\mathrm{B}$'s does not alter this picture. Fig.~\ref{screening}(c) shows the LDOS of two 
V$_\mathrm{B}$'s at different positions, with formation of two directional states oriented in different directions and not 
overlapping among each other. 

The formation of a moment is not a sufficient condition for long range magnetism~\cite{Droghetti_ASIC}. The picture emerging from 
our electronic structure calculations is that of magnetic moments associated to directional extended molecular orbitals at 
the B$_5$ clusters. We then expect magnetism only when the magnetic-moment-carrying levels of different B$_5$ clusters 
overlap. We have checked that distant V$_\mathrm{B}$'s are magnetically coupled by calculating the total energy difference 
between a ferromagnetic and an antiferromagnetic configuration of a $4\times2\times2$ supercell. We find that, when 
the two B$_5$ clusters are 8.23~\AA\ apart, there is an energy difference of 32~meV/vacancy ($\alpha=0.5$) in favor of the 
ferromagnetic solution. Although more calculations are needed at other V$_\mathrm{B}$ concentrations and for different 
V$_\mathrm{B}$-V$_\mathrm{B}$ separations, we may yet conclude that our calculations point to both the moment formation and 
exchange coupling between distant moments. A long range order is then expected when, typically randomly distributed, the magnetic 
moments form a percolation network. Two important considerations must be made here. First, as one missing B over the B$_6$ cluster 
is responsible for the moment and percolation is between the B$_6$ clusters, a cluster percolation
threshold of $x$ is achieved by removing only $x/6$ B ions. Second, we remark that the magnetism is related to the B$_6$ 
sub-lattice only, so that our results should be transferable to other XB$_6$ materials. However, we expect that the local 
electrical dipole associated to the Mg$^{2+}$ ions in cages bordered by $B_5$ clusters, will be different in direction and 
magnitude from that of the vacancy-free case.

Finally we wish to spend a few words on the possibility of the coexistence of both the ferromagnetic and anti-ferroelectric order. 
In general a ferroic state cannot be sustained in a metal because of the screening from the free electrons. The situation here 
is however more complex as, although there is density of states at the Fermi level, this is associated to partially localized 
states. Their screening can be inefficient to a point that local electric dipoles still form~\cite{Anderson,FEmetals}. Our results indeed 
confirm such a hypothesis as the Mg ions remain in their off-center positions as B vacancies are introduced in the cell. Notably, the 
strong directional nature of the charge density at the Fermi level suggests that such an inefficient screening persists locally even above 
the percolation threshold. 

In conclusion, we have investigated structural, electronic and magnetic properties of ideal and defective MgB$_{6-\delta}$ crystals. 
These present the coexistence of weak magnetism and potentially anti-ferroelectricity. The local magnetic moments originate
from a V$_\mathrm{B}$-related impurity band and the dipole moments from the displacement of the Mg$^{2+}$ ions within the 
cubic cage. Long-range magnetism can be sustained only above the magnetic percolation threshold and the anti-ferroelectric
state can coexist with a finite DOS at the Fermi level because of the incomplete screening from the localized charge. 
If proved experimentally, this will be the first example of coexistence of magnetic and electric order in a material not incorporating 
ions with partially occupied $d$ or $f$ shells.

This work is sponsored by Science Foundation of Ireland (SFI) under the CSET grant underpinning CRANN. NB acknowledges
support from SFI (grant No. 08/ERA/I1759). Computational resources have been provided by the HEA IITAC project managed by 
TCHPC. 


\end{document}